\documentclass[11pt,journal,onecolumn]{IEEEtran}

\usepackage{cite}
\usepackage{amsmath,amssymb,amsfonts}
\usepackage{graphicx}
\usepackage{textcomp}
\usepackage{xcolor}
\usepackage{algorithm}
\usepackage{algpseudocode}
\usepackage{array}
\usepackage{stfloats}
\usepackage{url}
\usepackage{verbatim}
\usepackage[nice]{nicefrac}
\usepackage{subcaption}
\usepackage{bbm}
\usepackage[colorlinks,bookmarksopen,bookmarksnumbered,citecolor=red,urlcolor=red]{hyperref}

\def\BibTeX{{\rm B\kern-.05em{\sc i\kern-.025em b}\kern-.08em
    T\kern-.1667em\lower.7ex\hbox{E}\kern-.125emX}}

\begin{document}

\title{MALRIS: \underline{Mal}icious Hardware in \underline{RIS}-Assisted Wireless Communications}

\author{Danish Mehmood Mughal, Daniyal Munir, Qazi Arbab Ahmed, Hans D. Schotten, \\Thorsten Jungeblut, Sang-Hyo Kim, and Min Young Chung
}




\maketitle

\begin{abstract}
Reconfigurable intelligent surfaces (RIS) enhance wireless communication by dynamically shaping the propagation environment, but their integration introduces hardware-level security risks. This paper presents the concept of Malicious RIS (MALRIS), where compromised components behave adversarially, even under passive operation. The focus of this work is on practical threats such as manufacturing time tampering, malicious firmware, and partial element control. Two representative attacks, power-splitting and element-splitting, are modeled to assess their impact. Simulations in a RIS-assisted system reveal that even a limited hardware compromise can significantly degrade performance metrics such as bit error rate, throughput, and secrecy metrics. By exposing this overlooked threat surface, this work aims to promote awareness and support secure, trustworthy RIS deployment in future wireless networks.
\end{abstract}


\begin{center}
    \textit{This is the author's version of the work accepted for presentation at IEEE CSCN 2025. \\
    © 2025 IEEE. Personal use of this material is permitted.}
\end{center}

\section{Introduction}
\label{sec:Intro}

Reconfigurable Intelligent Surfaces (RIS) are emerging as a cornerstone technology in the evolution of wireless networks, particularly in the context of 6G \cite{10596064}. By enabling the dynamic manipulation of the wireless propagation environment, RIS can significantly enhance signal strength, coverage, and energy efficiency through the passive reflection and phase control of electromagnetic waves \cite{9311936}. These capabilities position RIS as a key enabler of an intelligent, programmable, and highly efficient communication infrastructure \cite{9960899}.

In recent years, RIS has gained unprecedented attention from academia and industry (see \cite{10555049} and references therein) for its potential in future wireless networks. By enabling programmable manipulation of the wireless environment, RIS extends coverage in non-line-of-sight and shadowed areas, overcoming obstructions and dead zones \cite{10279515}. Through intelligent phase control, it enhances spectral efficiency, improving SINR and data rates \cite{10555049}. RIS operates passively without amplifying or generating new signals, offering greater energy efficiency than active repeaters or relays \cite{10268023}. It also mitigates interference and improves link reliability by dynamically steering beams based on real-time channel conditions \cite{10507199}. Additionally, RIS supports physical layer security by directing energy to intended receivers while limiting leakage to eavesdroppers \cite{10506802}. These features reduce latency, enhance reliability, and enable cost-effective network scaling, meeting key 6G performance targets.

Despite these promising capabilities, most RIS research has focused on performance gains via signal processing, channel estimation, and optimization~\cite{10555049}, assuming RISs are passive and trusted. However, as RISs are increasingly deployed in exposed environments, their potential misuse introduces new security risks~\cite{9112252}. Some studies have investigated malicious RIS configurations that threaten physical-layer security. For example,\cite{10143983} proposed a threat model where RISs assist eavesdroppers and introduced joint transmit power and RIS configuration optimization. In \cite{10693994}, benign and malicious RIS interactions were studied for MISO wiretap channels, using a game-theoretic max-min secrecy rate optimization under perfect CSI.
\begin{table*}[t]
\centering
\caption{Comparison of Recent Works on Malicious RIS Behavior}
\begin{tabular}{|p{0.7cm}|p{3.3cm}|p{3.3cm}|p{3.0cm}|p{5.0cm}|}
\hline
\textbf{Ref.} & \textbf{Focus} & \textbf{Attack Types} & \textbf{System Assumptions} & \textbf{Main Contributions} \\
\hline
\cite{10143983} & Eavesdropping enhancement using adversarial RIS & RIS-enhanced eavesdropping & Perfect CSI, Eve controls RIS & Introduces RIS as a tool for advanced eavesdropping under idealized assumptions \\
\hline
\cite{10693994} & Game-theoretic modeling of RIS behavior  & Passive jamming by adversarial RIS & Perfect CSI, discrete RIS phases, multiple RIS & Studies RIS interaction under benign and malicious roles using game theory \\
\hline
\cite{10516473} & Destructive phase-shift design by compromised RIS & SNR degradation via adversarial beamforming & Perfect CSI, \newline passive RIS & Shows effectiveness of malicious RIS beamforming under CSI uncertainty \\
\hline
\cite{11005403} & Broad taxonomy of control- and signal-level RIS threats & Jamming, pilot contamination, signal leakage, ND-RIS & Partial CSI, passive/active RIS   & Comprehensive survey and simulation of multi-modal RIS attack strategies \\
\hline
\cite{10856736} & ML-based detection of RIS threats during key generation & Adversarial RIS interference and phase manipulation & Partial CSI, adversarial RIS in key generation & Uses explainable ML for malicious RIS detection in key generation \\
\hline
This Work & Hardware-level RIS vulnerabilities & Power-splitting, Element-splitting & Perfect CSI,\newline passive RIS & Models partial hardware compromise of RIS, and quantify its physical-layer impact \\
\hline
\end{tabular}
\label{tab:malicious_RIS_comparison}
\end{table*}

Similarly,\cite{10516473} studied destructive beamforming by adversarial RISs and showed that phase misalignment can degrade secrecy capacity and throughput. More recently,\cite{11005403} provided a taxonomy of RIS-based attacks (e.g., jamming, eavesdropping, pilot contamination) and their impact in 6G, focusing on control-layer threats and adversarial optimization. Complementing these,~\cite{10856736} proposed an explainable adversarial learning framework to defend against malicious RIS behavior during physical-layer key generation. While promising, most works assume adversarial control at the configuration or algorithmic level.

While prior studies highlight RIS security, hardware-level vulnerabilities from physical deployment remain underexplored. Unlike conventional wireless nodes, RISs are installed on exposed surfaces without sensing or authentication, making them prone to tampering, malicious firmware, or element control during manufacturing. These risks are amplified by RIS-specific traits like passive operation and centralized control. Our work focuses on this underexplored attack surface by modeling low-level compromises, specifically, power-splitting and element-splitting strategies, and demonstrating their impact on system performance. The comparison in Table~\ref{tab:malicious_RIS_comparison} contextualize our approach relative to existing system-level and algorithm-level RIS threat models. By addressing threats that cannot be mitigated through signal processing alone, this work complements existing research and underscores the need for hardware-aware RIS security frameworks.

Our goal is to emphasize that RIS security must be considered from the hardware design level rather than added later. By highlighting the intersection of hardware security and emerging wireless architectures, this work seeks to spark broader dialogue on the safe and trustworthy deployment of RIS in future wireless networks. In summary, the following are our key contributions in this paper.

\begin{itemize}
    \item We introduce the concept of \textit{Malicious RIS} (MALRIS), where hardware-compromised RIS components covertly degrade communication or assist eavesdropping, even during passive operation.
    \item We identify and categorize practical hardware-level threats, including tampering, firmware manipulation, and electromagnetic interference, distinguishing them from traditional signal-level attacks.
    \item We propose two stealthy attack models, \textit{power-splitting} and \textit{element-splitting}, to represent partial RIS compromise, and analyze their impact on confidentiality, reliability, and availability.
    \item We evaluate these threats through simulations, showing their significant effect on bit error rate (BER), secrecy capacity, and outage probability, even with limited hardware manipulation.
\end{itemize}

The rest of the paper is structured as follows: 
Section~\ref{sec:HBTT} throws light on the hardware-based threats in RIS. Security implications of these attacks and risk assessment are presented in Section~\ref{sec:SIRA}. Impact of  MALRIS on the different performance metrics is detailed in Section~\ref{sec:Perf}. Finally, the paper concludes in Section~\ref{sec:Con} with some future directions.
\section{Hardware-based Threats to RIS-assisted communication }
\label{sec:HBTT}

RIS introduces unique vulnerabilities due to its reconfigurable hardware and reliance on external control circuits, creating new attack surfaces, especially under physical or electromagnetic access. This section outlines key hardware-level threats to RIS integrity and functionality. We consider a typical RIS-assisted communication scenario where a base station (BS) serves a user equipment (UE) via a RIS managed by an external controller (as in Fig. \ref{Fig1a_Basic}). An Eve nearby may exploit hardware vulnerabilities in the RIS or its control path. This highlights the hardware-centric attack surface inherent in RIS deployment, which is examined next.

\subsection{RIS Element Tampering}
RIS comprises dense arrays of programmable elements (e.g., FPGAs, PIN diodes, varactors, MEMS switches) to adjust wave phases \cite{10596064}. Often deployed on exposed structures (Fig. \ref{Fig1b_Tamper}), they are vulnerable to tampering, with MALRIS reflecting compromised signals to the UE. Assuming trusted design and manufacturing, physical attackers can still damage or rewire components, degrading beamforming and secrecy. As RIS evolves toward mmWave/THz 6G, miniaturization increases susceptibility to disruption \cite{10494372}. Even without physical access, high-power electromagnetic surges can induce faults remotely \cite{10759559}.
\subsection{Malicious Reconfiguration via Control Interface}
In addition to element tampering, RIS controllers are vulnerable as they manage phase shifts via commands from a central node over wired or wireless links. Without baseband processing or autonomous control, RIS primarily relies on control signal integrity. As shown in Fig.~\ref{Fig1c_CController}, the control interface can be a single point of failure if compromised. Here, we assume RIS controllers and elements are compromised during design or manufacturing, with no man-in-the-middle attacks. Malicious actors may exploit the supply chain by embedding hidden circuits or hardware Trojans in RIS components, triggered under specific conditions~\cite{10159361}. FPGA-based RIS can be reconfigured via illegitimate bitstreams~\cite{9474026}, while logic bombs may activate under certain RF conditions~\cite{Jin2016}. Without encrypted control channels, attackers can inject spoofed or replayed commands~\cite{ahmed-postconfig}. These hard-to-detect threats can cause coordination failures in multi-panel setups, making the control interface a key security bottleneck.
\subsection{Firmware Compromise and Backdoor}

Beyond element and control interface risks, RIS controllers face firmware-level threats. These controllers manage phase settings but usually lack security monitoring. Firmware may be compromised during manufacturing via logic bombs or hidden routines~\cite{9079514}, triggered by RF conditions, timing, or post-configuration~\cite{ahmed-postconfig}, altering phases or leaking data. As shown in Fig.~\ref{Fig1c_CController}, we consider trusted hardware with bitstreams maliciously modified during updates, physically or remotely. Unlike conventional nodes, RIS lacks user-facing software and runtime diagnostics, making firmware tampering highly stealthy. Insecure updates without cryptographic authentication~\cite{7827620} let attackers install persistent malware for surveillance or denial-of-service. RIS’s passive design, limited processing, and exposed deployment amplify these risks, making firmware-level threats especially critical.
\subsection{Side-Channel Attacks}
\begin{figure}[t]
   \centering
   \subfloat[]{\includegraphics[width=0.25\linewidth]{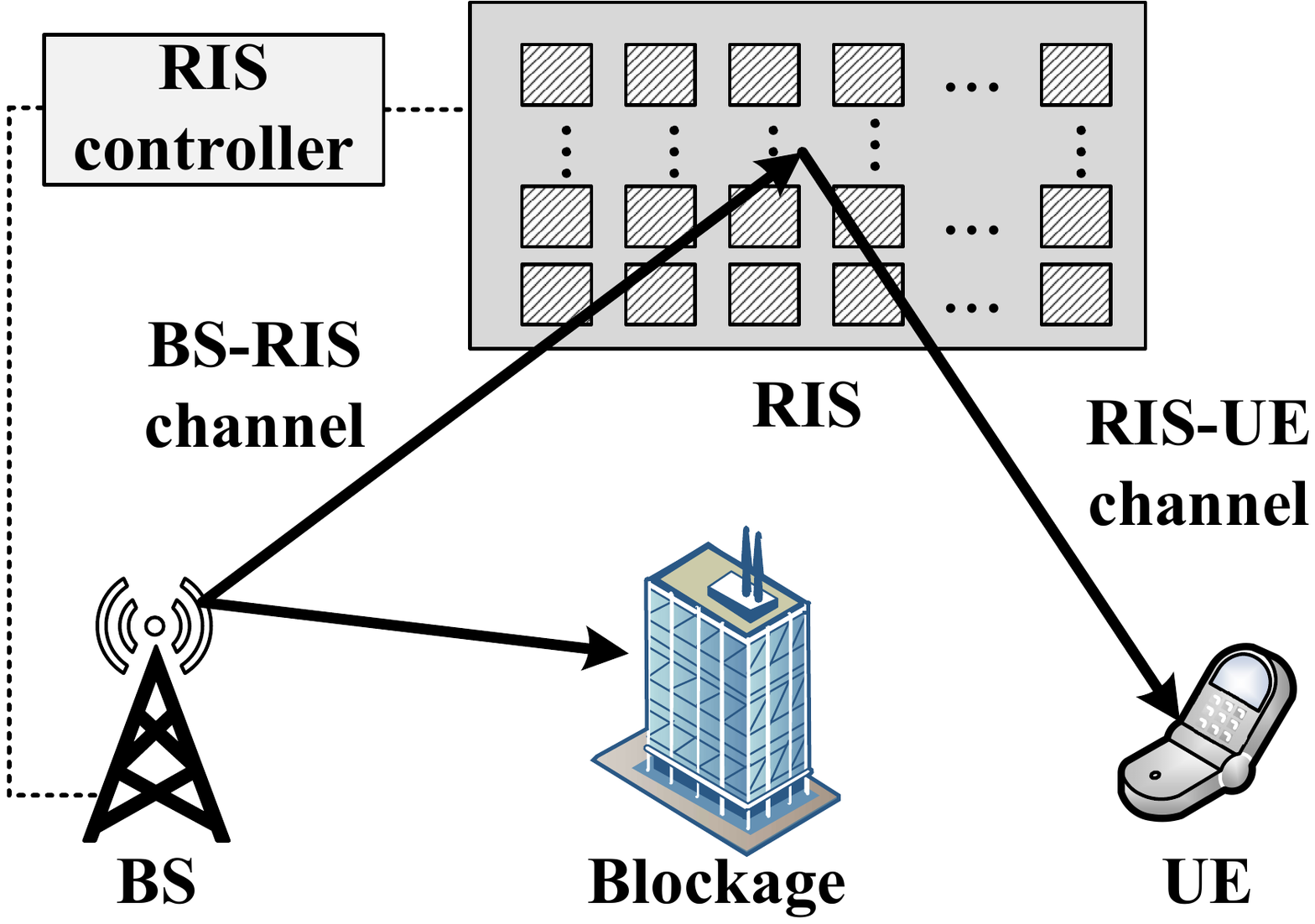}\label{Fig1a_Basic}}
   \hfill
   \subfloat[]{\includegraphics[width=0.25\linewidth]{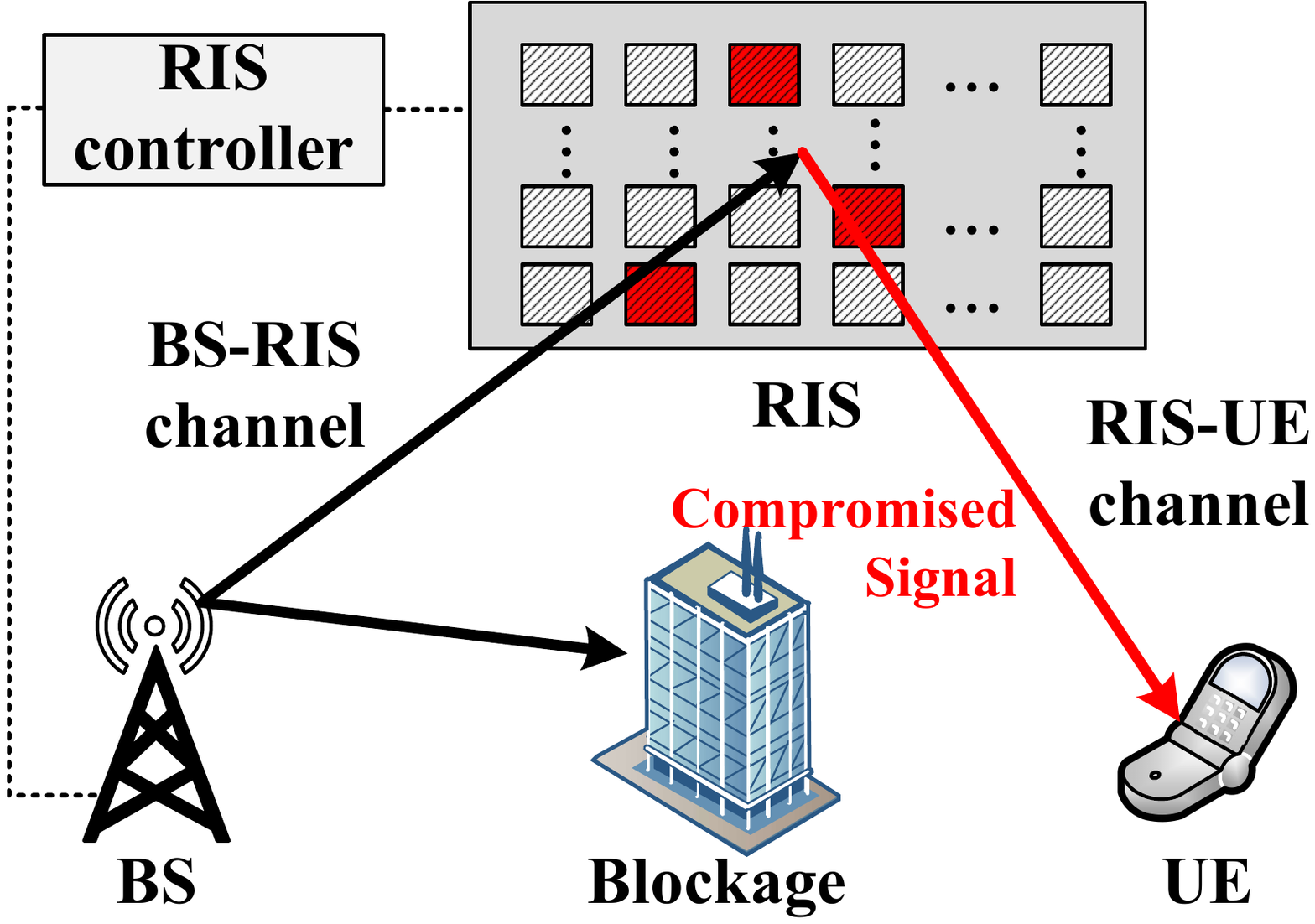}\label{Fig1b_Tamper}}
   \hfill
   \subfloat[]{\includegraphics[width=0.25\linewidth]{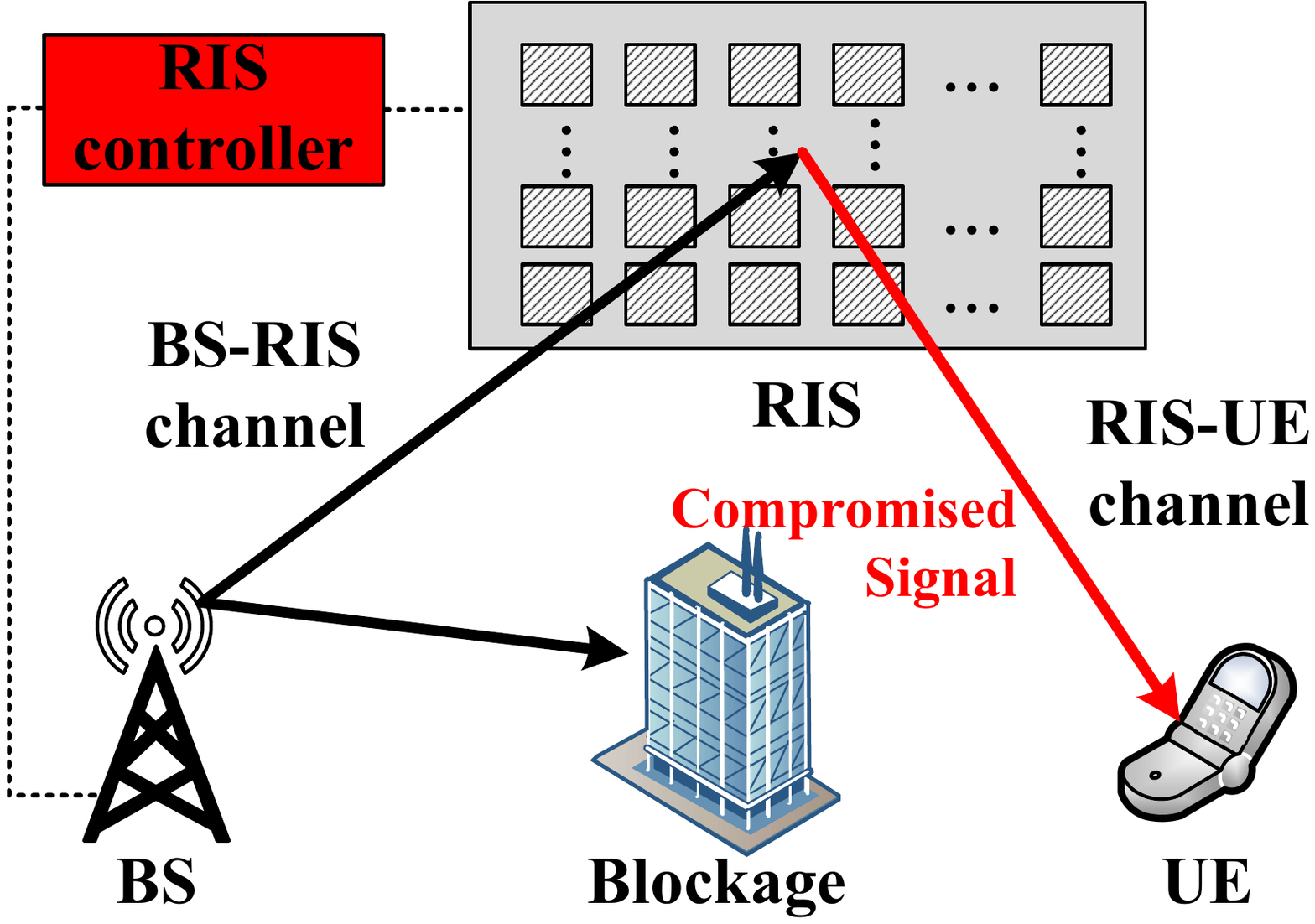}\label{Fig1c_CController}}
   \hfill
   \subfloat[]{\includegraphics[width=0.25\linewidth]{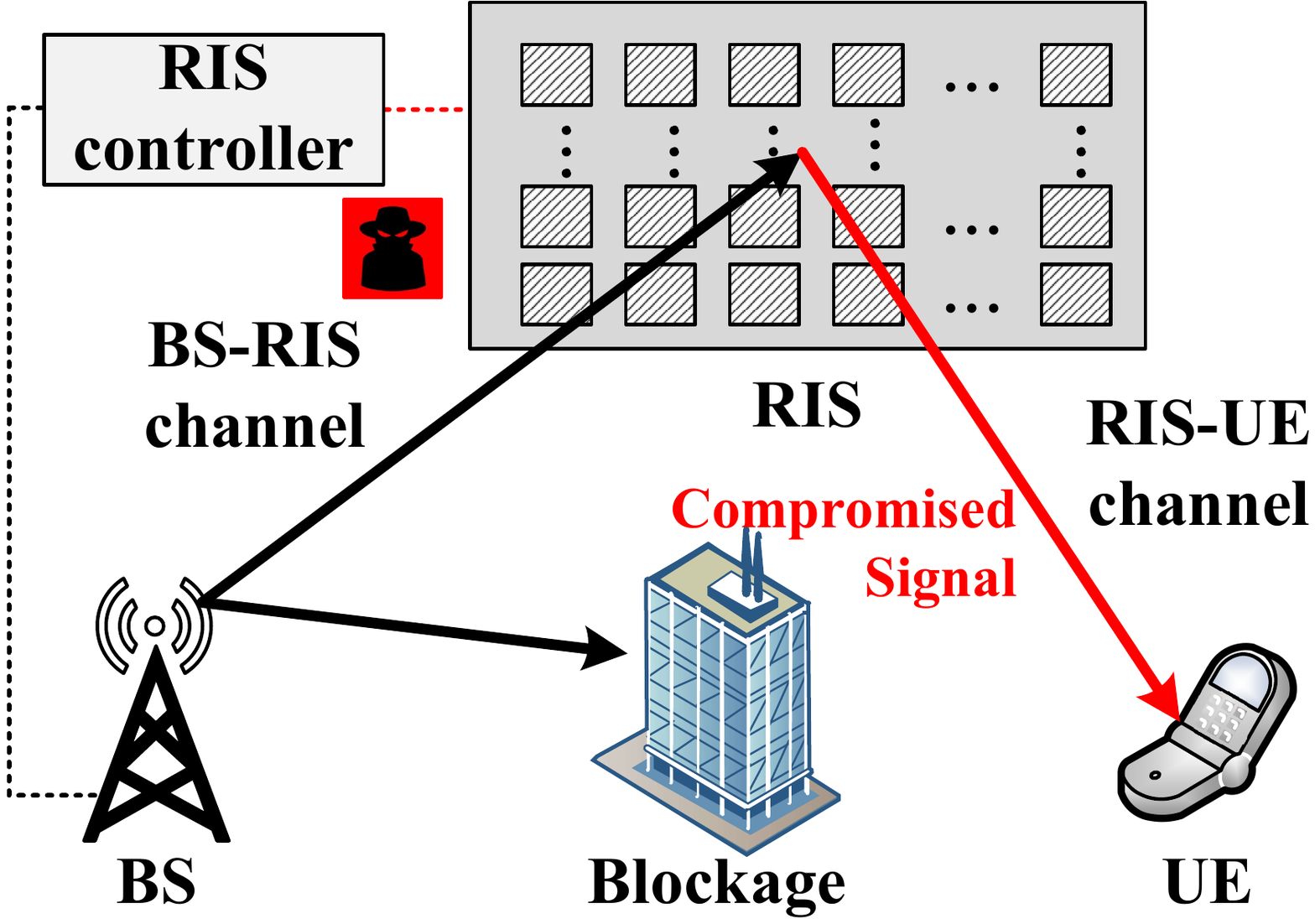}\label{Fig1d_Sidelink}}
   \caption{(a) RIS-assisted network mode (b) RIS element tampering (c) Compromised RIS controller (d) Side-channel attack.}
   \label{fig1:main}
\end{figure}

Even without direct access, attackers can exploit side-channel leaks, e.g., electromagnetic radiation, power fluctuations~\cite{8342177}, or acoustic signals (as in Fig.~\ref{Fig1d_Sidelink}), from RIS control circuitry, revealing phase settings, beam directions, or user activity. Here, we assume trusted design and manufacturing, but the channel between the RIS controller and the RIS is compromised by man-in-the-middle attacks. An attacker can passively observe side-channel patterns to infer operations, correlate data, or fingerprint RIS panels and firmware, without system contact. These stealthy attacks need no modifications, use off-the-shelf tools, leave no trace, and are hard to defend against due to RIS’s passive design.
\section{Security Implications and Risk Assessment}
\label{sec:SIRA}
Hardware-level attacks on RIS components pose serious risks to wireless system security. This section analyzes these threats using the Confidentiality, Integrity, and Availability (CIA) framework to highlight how they can enable unauthorized access, disrupt services, or degrade performance in real-world deployments.
\subsection{Confidentiality Risks}
Confidentiality in RIS-assisted networks is especially vulnerable due to their passive nature and reliance on external control. Adversaries can exploit several threats to intercept sensitive information silently:
\textit{a) Malicious reconfiguration} can steer signals toward nearby eavesdroppers. \textit{ b) Backdoor insertion}, such as a hardware Trojan, during device manufacturing, may leak configuration or channel data. \textit{c) Side-channel leakage} reveals beam directions or user movement through emissions.
These attacks bypass traditional cryptographic defenses by targeting the physical layer, enabling undetectable surveillance and undermining the spatial privacy that directional systems are meant to ensure.
\subsection{Integrity Risks}
Integrity in RIS-assisted systems is particularly fragile due to their centralized, passive control. Adversaries can manipulate RIS behavior to cause misleading or degraded operations through:
\textit{a) element tampering} involves physically altering reflectors to misdirect beams or degrade precision,
\textit{b) firmware compromise} by injecting malicious logic to trigger unauthorized phase changes or noise only during operation, and 
\textit{c) spoofed commands} to mislead the RIS into applying attacker-defined configurations. 
Such attacks may go undetected yet significantly disrupt communication. In cooperative or high-precision systems, even subtle integrity breaches can cascade, destabilizing the entire transmission process.
\subsection{Availability Risks}
Availability attacks on RIS aim to disrupt service or disable components, and are especially dangerous due to RIS’s passive design and dependence on precise control. Key threats include: 
\textit{a) Physical tampering} or EM interference can damage or desynchronize RIS elements, degrading coverage.
\textit{c) Unauthorized commands} may overload the system with invalid reconfigurations, disabling functionality.
\textit{d) Malicious firmware} updates during operation can “brick” the RIS, with no built-in recovery.
\textit{e) Jamming amplification}, where compromised hardware in RIS reflects and enhances jamming signals, extends the attack’s reach and impact.
These threats can cripple systems, especially in cooperative MIMO or URLLC, while escaping detection due to minimal RIS monitoring. What was designed to enhance performance can become a tool for disruption. 
\section{Performance Evaluation and Discussions}
\label{sec:Perf}
This section presents the system model and evaluates the performance of MALRIS. We examine two scenarios where MALRIS redirects signals to Eve: by manipulating a subset of elements or diverting part of the signal power.
 \begin{figure}[t]
   \centering
   \subfloat{\includegraphics[width=0.45\linewidth]{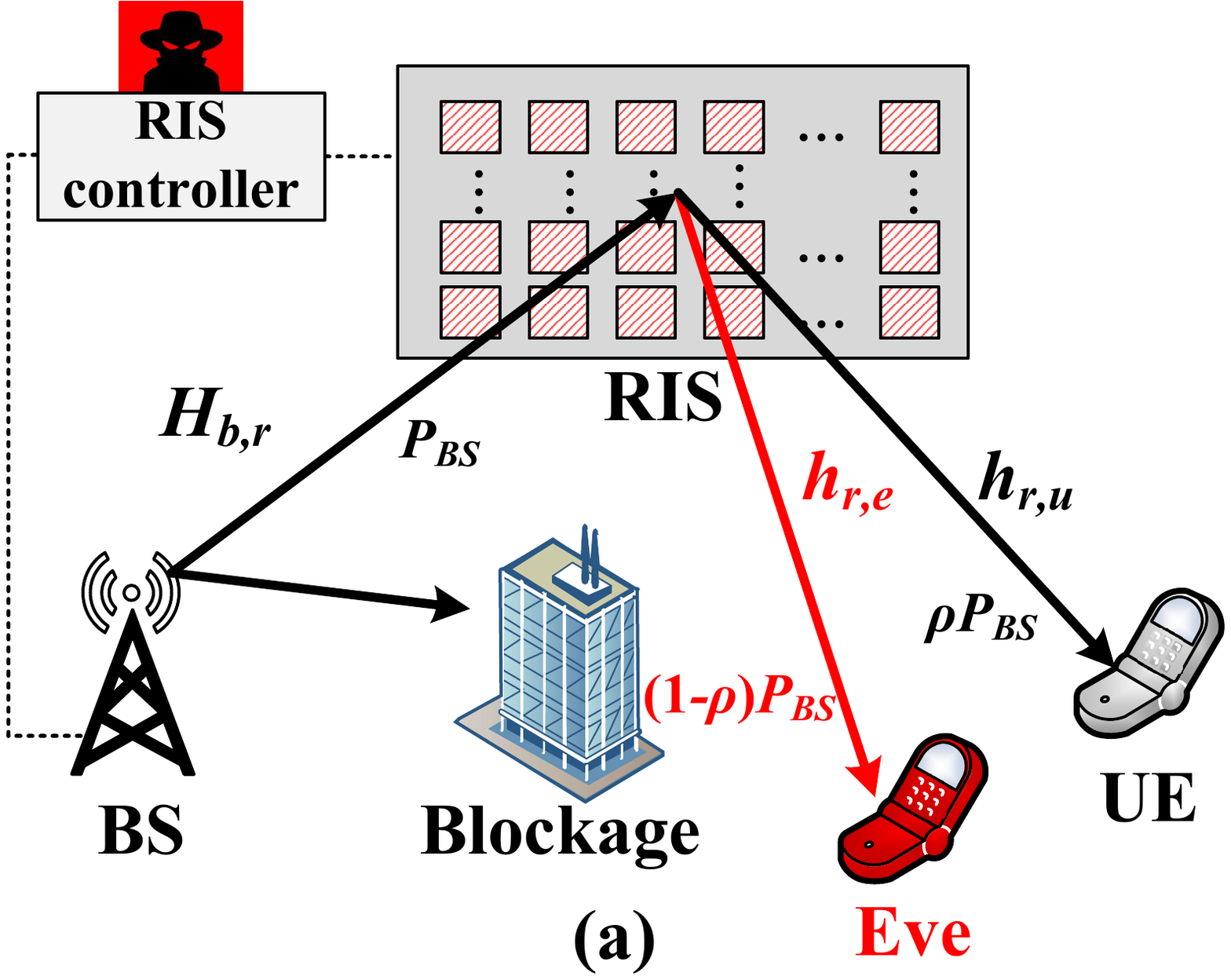}
   \label{Fig2a_PowerSplit}}
   \hfill
   \subfloat{\includegraphics[width=0.45\linewidth]{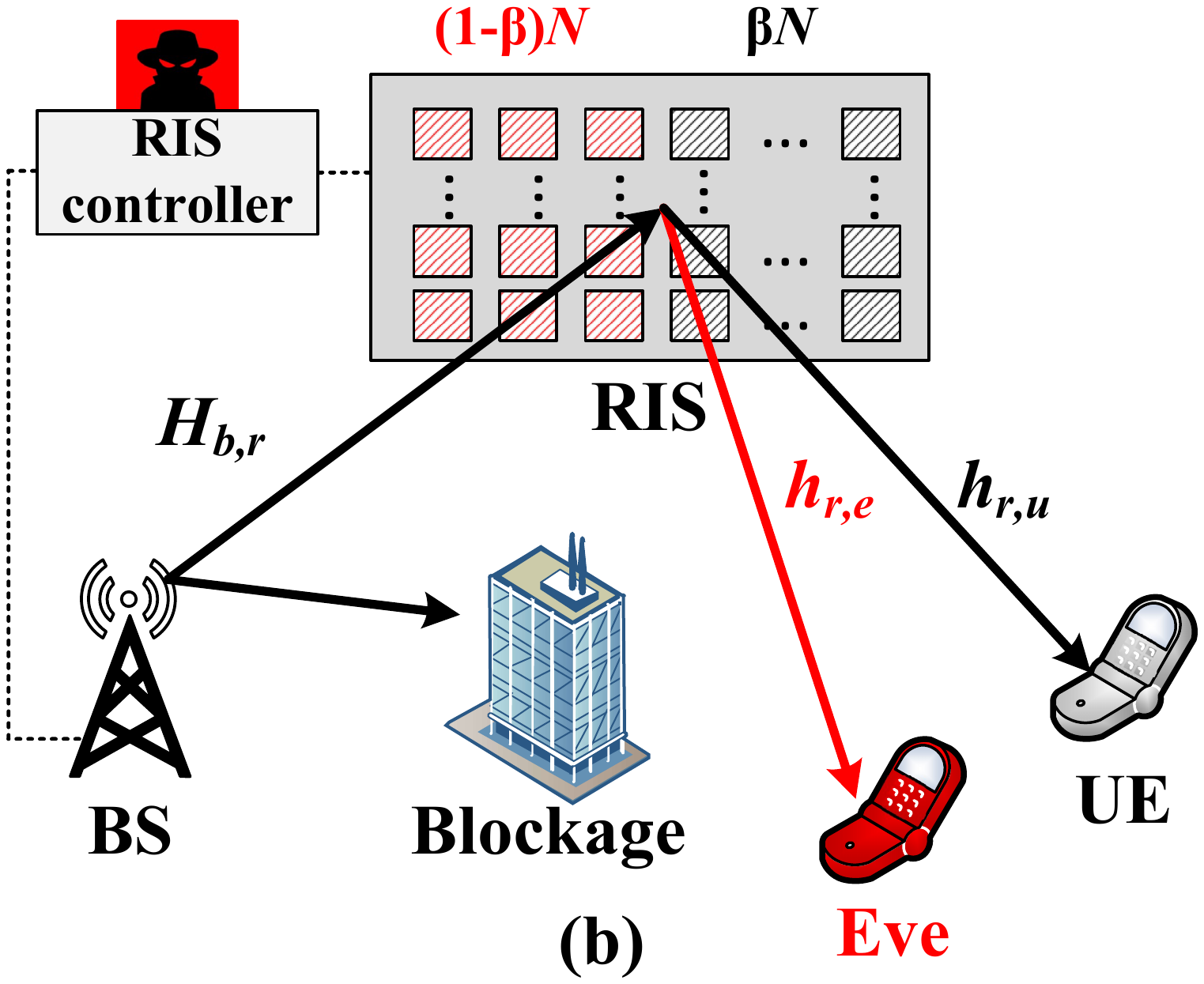}\label{Fig2b_Ant_Split}}
   \caption{RIS-assisted communication under MALRIS threats, illustrating signal leakage via malicious reflection strategies: (a) power splitting and (b) element splitting.}
   \vspace{-4mm}
   \label{fig:main}
 \end{figure}
\subsection{System Model}
We consider a RIS-assisted wireless system where a BS with $M$ antennas serves a single-antenna UE via a RIS with $N$ elements. Perfect CSI for the BS-RIS-UE path is assumed. The RIS is malicious, which means MALRIS operates normally, assisting BS-to-UE communication. However, when a nearby Eve is present, it maliciously redirects part of the BS signal toward Eve.
\subsection*{1. Channel Modeling}
Let the channel from BS to RIS be denoted as $\mathbf{H}_{\text{b,r}} \in \mathbb{C}^{N \times M}$, and the channel from RIS to UE be $\mathbf{h}_{\text{r,u}} \in \mathbb{C}^{1 \times N}$. We assume that each channel follows a Rician fading model such that,
\begin{equation}
\mathbf{H}_{\text{b,r}} = \sqrt{PL_{\text{b,r}}} \left( \sqrt{\frac{\kappa}{\kappa+1}} \mathbf{H}^{\text{LOS}}_{\text{b,r}} + \sqrt{\frac{1}{\kappa+1}} \mathbf{H}^{\text{NLOS}}_{\text{b,r}} \right),
\end{equation}
\begin{equation}
\mathbf{h}_{\text{r,u}} = \sqrt{PL_{\text{r,u}}} \left( \sqrt{\frac{\kappa}{\kappa+1}} \mathbf{h}^{\text{LOS}}_{\text{r,u}} + \sqrt{\frac{1}{\kappa+1}} \mathbf{h}^{\text{NLOS}}_{\text{r,u}} \right),
\end{equation}
where $\kappa$ is the Rician factor, and $PL_{\text{b,r}}$ and $PL_{\text{r,u}}$ denote large-scale path loss terms modeled as
$PL = d^{-\alpha}$, with $d$ being the distance between two network entities and $\alpha$ is the path-loss exponent. Similarly, the channel from RIS to Eve is denoted by $\mathbf{h}_{\text{r,e}} \in \mathbb{C}^{1 \times N}$, also modeled via Rician fading and subject to large-scale fading $PL_{\text{r,e}}$.
\subsection{Signal Model Without MALRIS}
Let the BS transmit signal $s$ with power $P_{\text{BS}}$ using beamforming vector $\mathbf{x} \in \mathbb{C}^{M \times 1}$ satisfying $\|\mathbf{x}\|^2 = P_{\text{BS}}$. The beamforming vector $\mathbf{x}$ is designed using Maximum Ratio Transmission (MRT) based on the effective cascaded channel. Assuming perfect CSI of the BS--RIS--UE link, the optimal beamforming direction is given by:
\begin{equation}
    \mathbf{x} = \sqrt{P_{\text{BS}}} \cdot \frac{\left( \mathbf{H}_{\text{b,r}}^{\dagger} \boldsymbol{\Theta}^{H} \mathbf{h}_{\text{r,u}}^{H} \right)}{\left\| \mathbf{H}_{\text{b,r}}^{\dagger} \boldsymbol{\Theta}^{H} \mathbf{h}_{\text{r,u}}^{H} \right\|},
\end{equation}
where $\dagger$ denotes the Hermitian (conjugate transpose), and $\left\|\cdot\right\|$ is Frobenius norm. This ensures the BS transmits along the strongest direction of the cascaded channel $\mathbf{h}_{\text{r,u}} \boldsymbol{\Theta} \mathbf{H}_{\text{b,r}}$, maximizing the received power at the UE.
Let $\boldsymbol{\Theta} = \mathrm{diag}(e^{j\theta_1}, ..., e^{j\theta_N})$ be the RIS phase shift matrix optimized using perfect CSI of the BS-RIS-UE link. The received signal at the UE is given by:
\begin{equation}
y_{\text{u}} = \mathbf{h}_{\text{r,u}} \boldsymbol{\Theta} \mathbf{H}_{\text{b,r}} \mathbf{x} s + n_{\text{u}},
\end{equation}
where $n_{\text{u}} \sim \mathcal{CN}(0, \sigma_u^2)$ is the additive white Gaussian noise. $ s \sim \!\mathcal{CN}(0, 1)$ is a transmitted symbol represented by zero-mean circularly symmetric complex Gaussian random variables with unit power satisfying $\mathbb{E}[|s|^2] = 1$.
The effective end-to-end channel gain is:
$h_{\text{eff}}^{(u)} = \mathbf{h}_{\text{r,u}} \boldsymbol{\Theta} \mathbf{H}_{\text{b,r}} \mathbf{x},$
and the resulting SNR $(\gamma_{\text{u}})$ at UE is:
\begin{equation}
\gamma_{\text{u}} = \frac{|h_{\text{eff}}^{(u)}|^2}{\sigma_u^2} = \frac{|\mathbf{h}_{\text{r,u}} \boldsymbol{\Theta} \mathbf{H}_{\text{b,r}} \mathbf{x}|^2}{\sigma_u^2}.
\end{equation}
\subsection{Signal Model During MALRIS Behavior}
We assume that the MALRIS starts the security attack after $T/2$ time slots ($T$ is the total number of slots), whereby it splits its reflection resources between the UE and Eve. In this paper, we consider two scenarios: power splitting and RIS element splitting.

\textbf{(a) Power-based Splitting:} We studied a possible attack whereby MALRIS splits incoming signal power such that $\rho P_{\text{BS}}$ is directed toward the UE and $(1 - \rho) P_{\text{BS}}$ is reflected toward the Eve, with $\rho \in [0,1]$ is the power splitting factor. 
Specifically, signal power can be manipulated by attack vectors such as malicious reconfiguration via control interface. In this scenario, effective channel gain and SNR $(\gamma_{\text{u}})$ at the UE are given as
\begin{equation}
h_{\text{eff}}^{(u)} = \sqrt{\rho} \cdot \mathbf{h}_{\text{r,u}} \boldsymbol{\Theta} \mathbf{H}_{\text{b,r}} \mathbf{x},
\end{equation}
\begin{equation}
\gamma_{\text{u}} = \frac{|h_{\text{eff}}^{(u)}|^2}{\sigma_u^2}.
\end{equation}
For Eve, the received signal is modeled statistically as $y_{\text{e}} = \sqrt{1 - \rho} \cdot z + n_{\text{e}}$, with the resulting SNR $(\gamma_{\text{e}})$ as
\begin{equation}
\gamma_{\text{e}} = \frac{(1 - \rho)|z|^2}{\sigma_e^2}.
\end{equation}
Here, $z$ is an unknown leakage signal and $\gamma_{\text{e}}$ is approximated and simulated empirically.

\textbf{(b) Element-based Splitting:} 
In addition to power-based splitting, MALRIS can manipulate a subset of RIS elements to facilitate spoofing or side-channel attacks. Let $\beta \in [0,1]$ represent the fraction of RIS elements allocated to the UE. The remaining fraction $(1-\beta)$ is used by MALRIS to reflect the signal to Eve. The MALRIS phase matrix can be partitioned accordingly, and the effective channel for UE in this scenario is computed as:
\begin{equation}
h_{\text{eff}}^{(u)} = \mathbf{h}_{\text{r,u}}^{(\beta)} \boldsymbol{\Theta}_{\text{u}} \mathbf{H}_{\text{b,r}}^{(\beta)} \mathbf{x},
\end{equation}
\begin{equation}
\gamma_{\text{u}} = \frac{|h_{\text{eff}}^{(u)}|^2}{\sigma_u^2}.
\end{equation}
Here, $\mathbf{H}_{\text{b,r}}^{(\beta)}$ is the subset of $\mathbf{H}_{\text{b,r}}$ matrix, representing the channels for RIS-UE communication. Since the CSI of the RIS-Eve link is unknown, the effective signal at Eve is modeled as $y_{\text{e}} = z + n_{\text{e}}$, and the resultant SNR at Eve $\gamma_{\text{e}}$ is
\begin{equation}
\gamma_{\text{e}} = \frac{|z|^2}{\sigma_e^2},
\end{equation}
where $z$ is a random leakage signal dependent on $(1-\beta)N$ RIS elements, and $n_{\text{e}} \sim \mathcal{CN}(0, \sigma_{e}^2)$ is noise with power $\sigma_{e}^2$.
\subsection{Performance Evaluation}
We evaluate throughput, secrecy capacity, secrecy outage probability, and (BER) as performance metrics for the proposed model under two scenarios: with and without a compromised RIS. For the first half of the simulation time, i.e., during $T/2$ time slots (where $T$ is the total simulation time), the RIS operates normally. After $T/2$, the RIS becomes malicious and begins splitting either its antenna elements or the transmitted power between the legitimate user and the Eve. Throughput is computed in terms of Shannon capacity as
\begin{equation}
C = \log_2(1 + \gamma), \quad \text{[bps/Hz]},
\end{equation}
and secrecy capacity, defined as the maximum rate at which secure communication can be achieved  \cite{4035982}, given as:
\begin{equation}
C_s = \max\left(0, \log_2(1 + \gamma_{\text{u}}) - \log_2(1 + \gamma_{\text{e}})\right).
\end{equation}

Moreover, we computed secrecy outage probability ($P_{\text{out}}$), which is the probability that the secrecy  capacity is less than the target secrecy rate ($R_s$), given as
$P_{\text{out}} = \Pr(C_s < R_s)$.
We have estimated $P_{\text{out}}$ via Monte Carlo simulation for performance evaluation over multiple channel realizations as:
\begin{equation}
P_{\text{out}} \approx \frac{1}{N_{\text{sim}}} \sum_{i=1}^{N_{\text{sim}}} \mathbbm{1}\cdot({ C_s^{(i)} < R_s }),
\end{equation}
where $\mathbbm{1}$ is the indicator function.

Lastly, to compute the BER at UE, we modulated random binary data using QPSK (where two bits are mapped per symbol), transmitted through the channel, and demodulated at the receiver. The BER is computed as the ratio of incorrectly decoded bits to the total number of transmitted bits.
\begin{table}[t]
\centering
\caption{Simulation Parameters}
\label{tab:Param_Val}
\begin{tabular}{|l|l||l|l|}
\hline
\textbf{Parameter}       & \textbf{Value} & \textbf{Parameter}                    & \textbf{Value}\\ 
\hline 
\hline
BS location              & (0,0)          & RIS location                          & (50,20)      \\ 
\hline 
UE location              & (75,0)         & Eve location                          & (50,-20)     \\ 
\hline
BS power ($P_{BS}$) (db) & 10, 20         & RIS elements ($N$)                    & 32, 64       \\ 
\hline
BS antenna ($M$)         & 4              & Noise power $\sigma_u^2$, $\sigma_e^2$ & $10^{-7}$   \\
\hline
Rician factor ($\kappa$) & 5              & Path loss exponent ($\alpha$)         & 3            \\ 
\hline
Simulation time ($T$)    & 50             & Target secrecy rate ($R_s$)           & 1 bps/Hz      \\ 
\hline
\end{tabular}
\vspace{-5mm}
\end{table}
\begin{figure}[b]
\vspace{-7mm}
  \centering
  \subfloat[]{\includegraphics[width=0.48\linewidth]{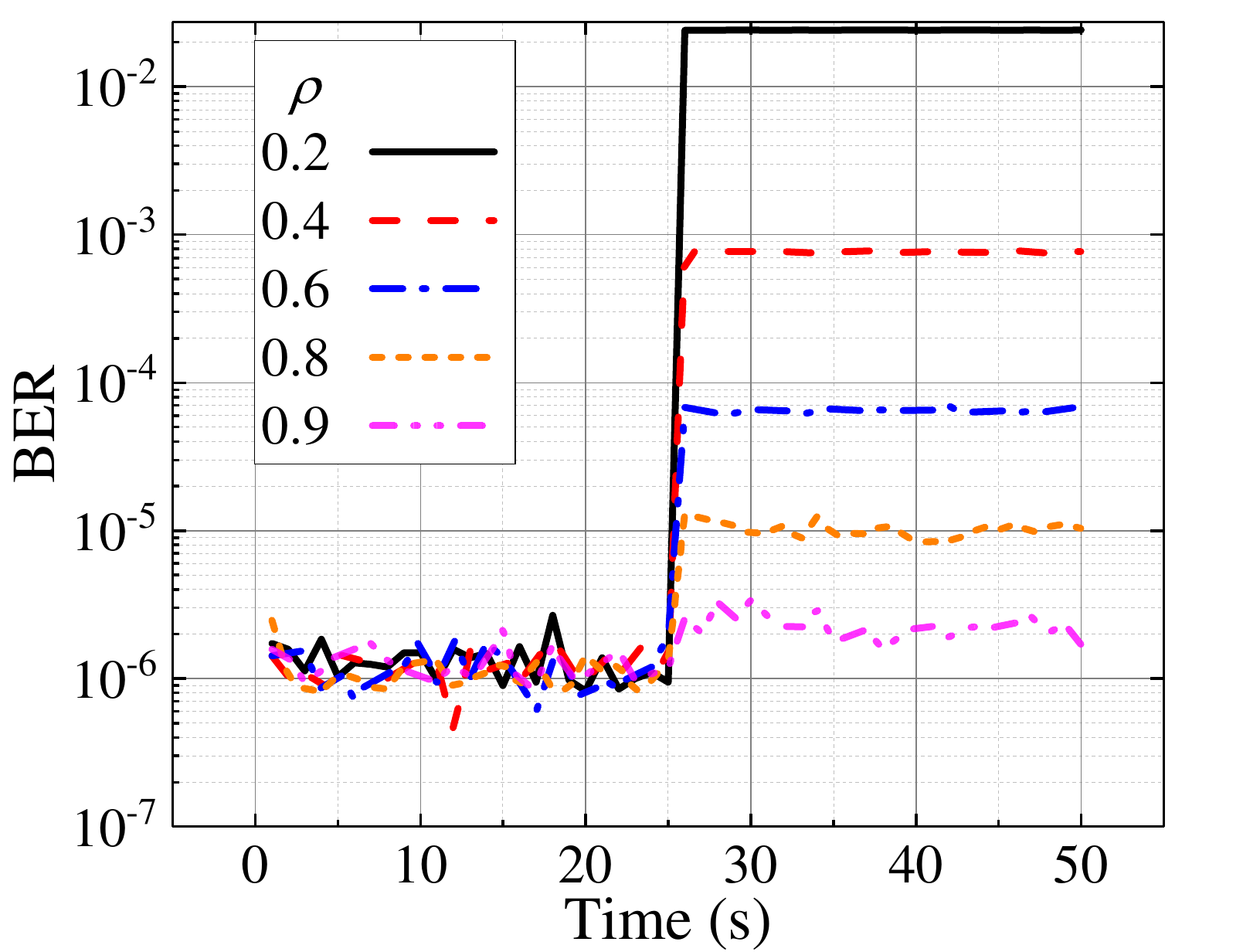}
  \label{fig:Fig3_Beta_BER}}
  \hfill
  \subfloat[]{\includegraphics[width=0.48\linewidth]{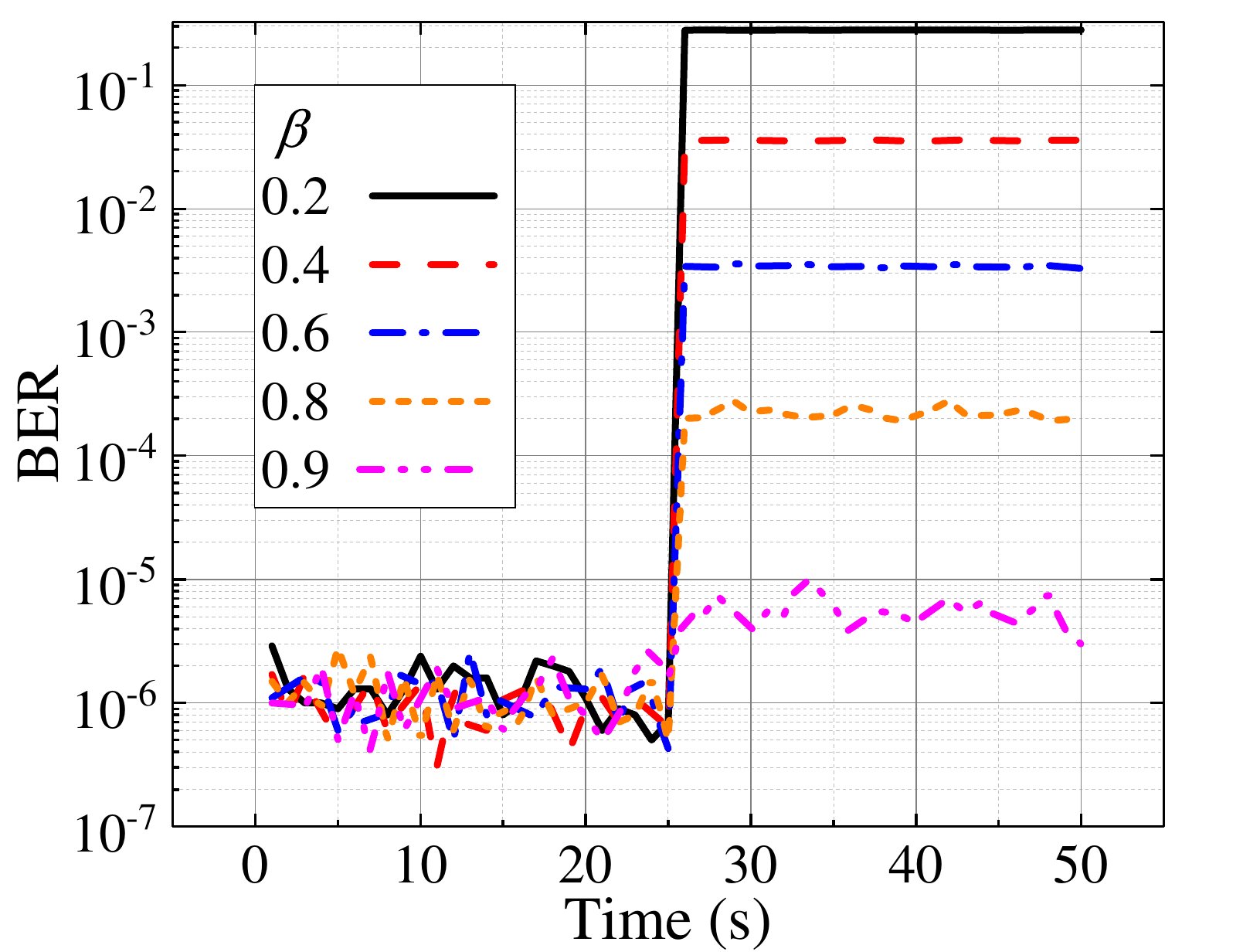}
  \label{fig:Fig3_Beta_TH}}
  \caption{BER performance of MALRIS-assisted network under (a) power splitting and (b) antenna element-splitting attacks.}
  \label{fig3:AntSplitPerf}
\end{figure}

The security vulnerabilities when Eve manipulates some of the antenna elements and power splitting for malicious activities are presented in terms of BER in Fig. \ref{fig3:AntSplitPerf}. General parameter settings for these results are summarized in Table \ref{tab:Param_Val}.
As illustrated in Figs. \ref{fig:Fig3_Beta_BER} and \ref{fig:Fig3_Beta_TH}, the BER increases sharply when a malicious attack is launched by MALRIS. Intuitively, for lower values of $\rho$, less power is allocated for the UE, resulting in degraded BER due to weak signal strength, as shown in Fig. \ref{fig:Fig3_Beta_BER}. In contrast, higher $\rho$ values result in reflecting the signal to the UE with more power, improving the BER and resulting SNR.
Fig. \ref{fig:Fig3_Beta_TH} further demonstrates the impact of the antenna element splitting on BER performance during a MALRIS attack. Lower $\beta$ values result in fewer RIS elements allocation for the UE and degraded BER performance due to weaker signal strength. Comparing the two figures, it is evident that antenna splitting results in worse BER performance than power splitting. This degradation is primarily attributed to the significant reduction in the RIS gain available to the UE during antenna splitting.

\begin{figure}[t]
  \centering
  \subfloat[]{\includegraphics[width=0.5\linewidth]{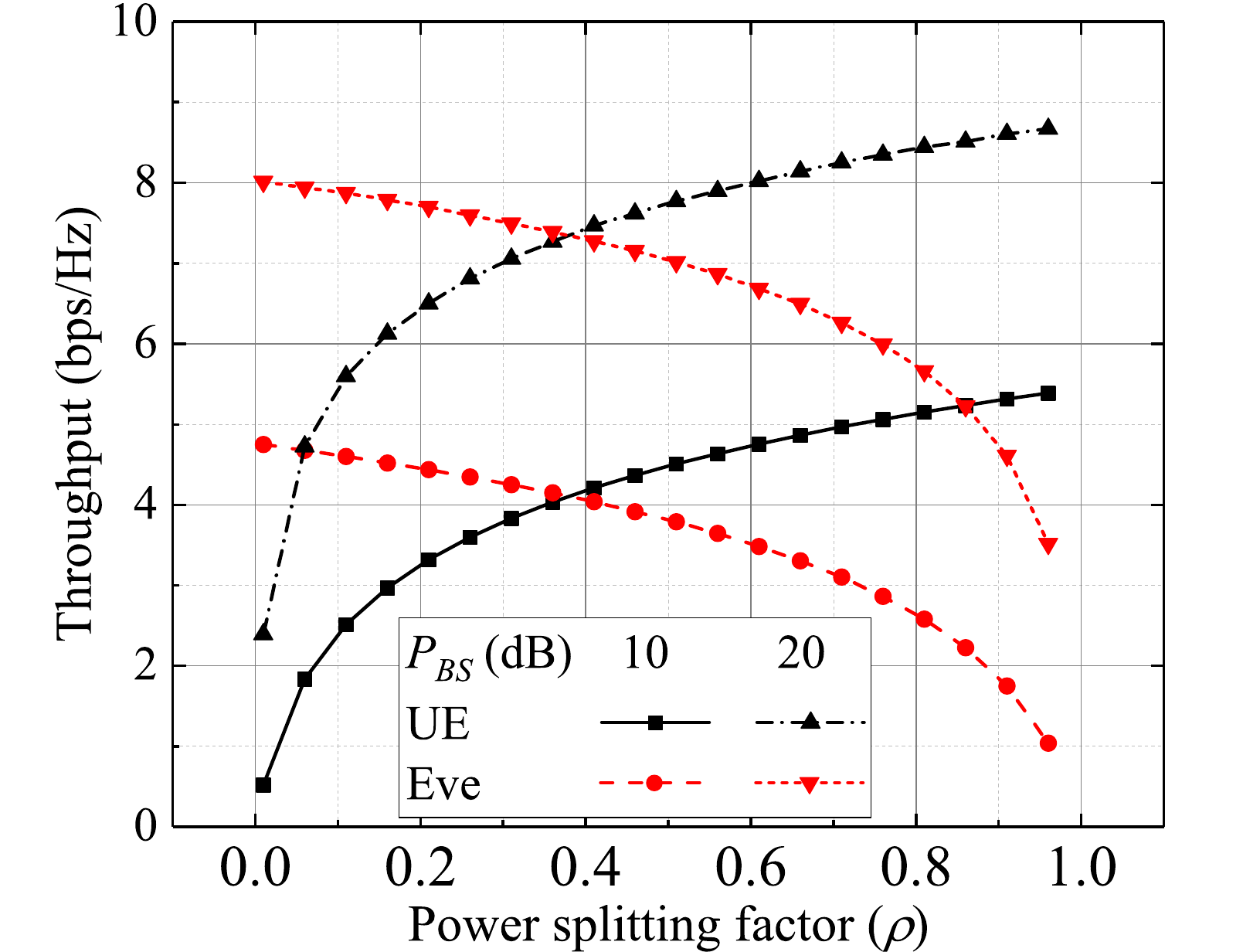}\label{fig:Fig5_Rho}}
  \hfill
  \subfloat[]{\includegraphics[width=0.5\linewidth]{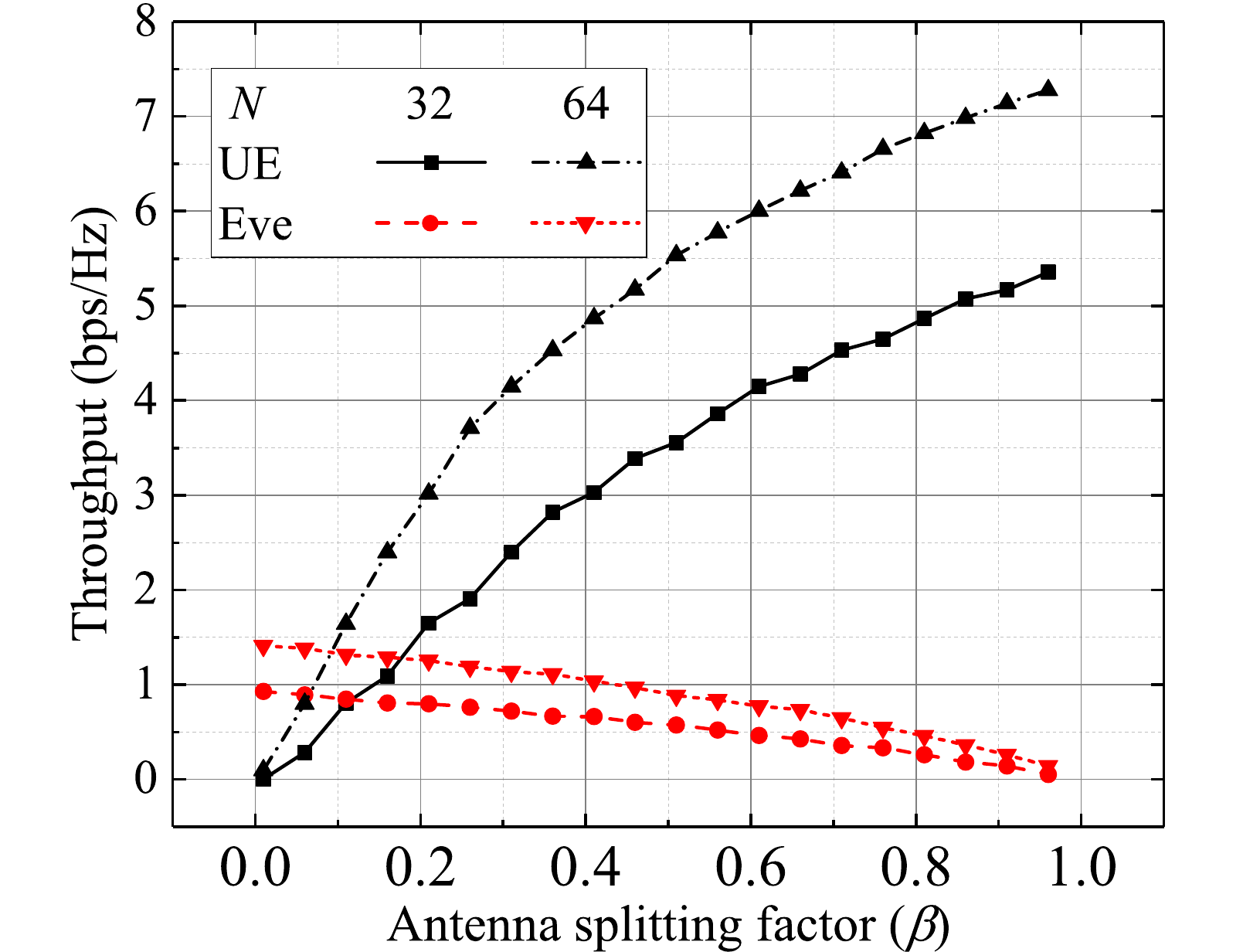}\label{fig:Fig5_Beta}}
  \caption{Throughput of the MALRIS-assisted network under (a) power-splitting and (b) antenna element-splitting attacks.}
  \label{fig:Throughput}
\vspace{-5mm}
\end{figure}

Fig.~\ref{fig:Throughput} shows the throughput of the UE and Eve as functions of $\rho$ and $\beta$. In Fig.~\ref{fig:Fig5_Rho}, throughput is plotted against $\rho$ for different $P_{BS}$ values. When $\rho$ is small, most of the power is directed to Eve, resulting in high Eve throughput and low UE throughput. As $\rho$ increases, UE throughput improves while Eve's decreases. Higher $P_{BS}$ leads to greater throughput in both cases owing to the higher signal strength at the receiving entities. However, at mid-range $\rho$ values, UE and Eve throughput become comparable, raising security concerns. Fig.~\ref{fig:Fig5_Beta} illustrates throughput versus the antenna splitting factor $\beta$. Increasing $\beta$ allocates more elements to the UE, enhancing its throughput. Hereby, as Eve is farther from RIS compared to UE, its performance is mostly lower, when $\beta$ increases as the RIS gain becomes minimal. The RIS gain is noticeable for the UE case, whereby the difference between  $N=32$ and $N=64$ increases as the value of $\beta$ increases.

\begin{figure}[t]
  \centering
  \subfloat[]{\includegraphics[width=0.5\linewidth]{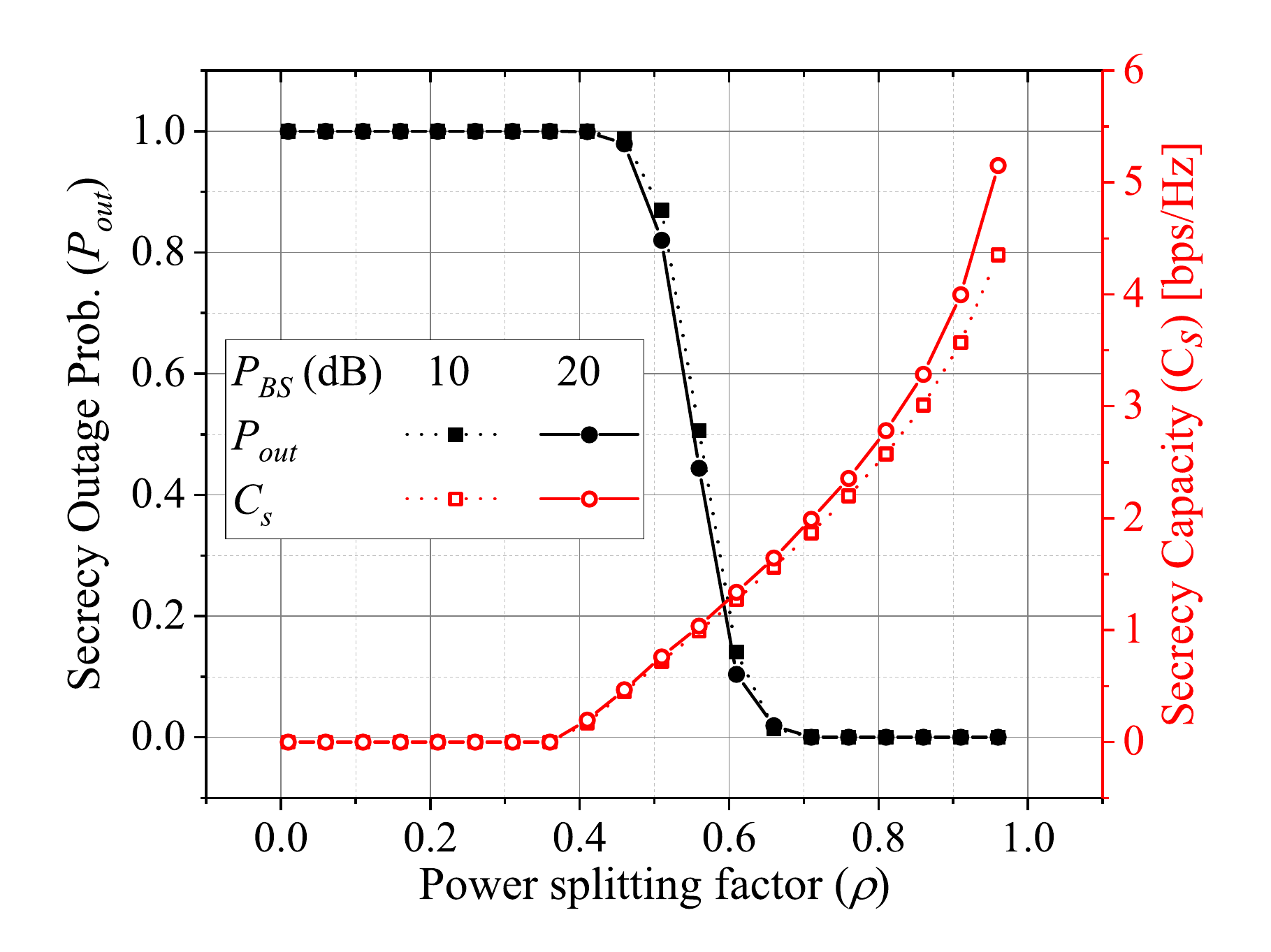}\label{fig:Fig5_Rho_Sec}}
  \hfill
  \subfloat[]{\includegraphics[width=0.5\linewidth]{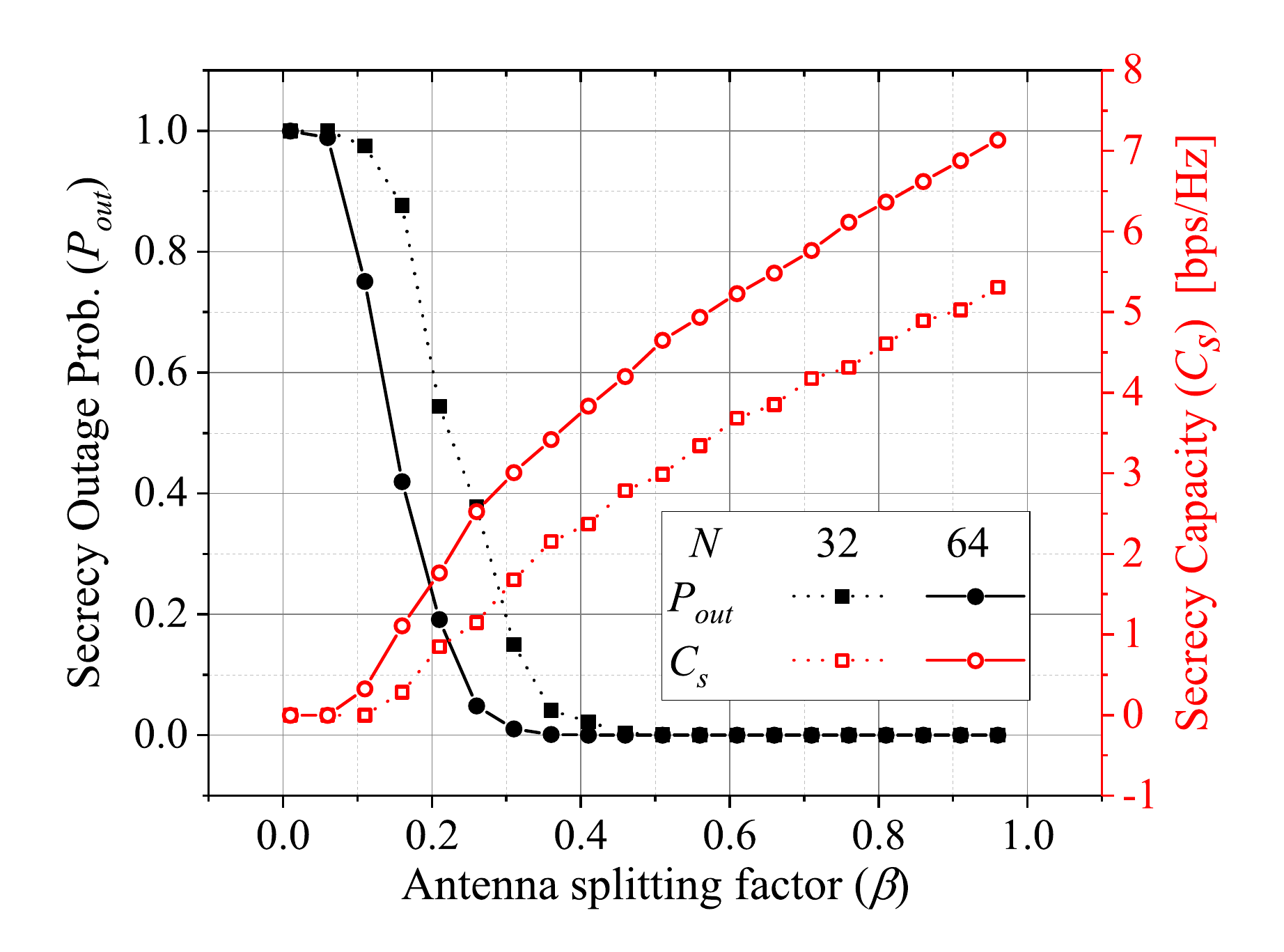}\label{fig:Fig5_Beta_Sec}}
  \caption{Secrecy outage probability and capacity of the MALRIS-assisted network under (a) power-splitting and (b) antenna element-splitting attacks.}
  \label{fig:SecOut}
  \vspace{-5mm}
\end{figure}

Fig.~\ref{fig:SecOut} illustrates the secrecy outage probability ($P_{out}$) and secrecy capacity ($C_s$) of the UE under power and antenna splitting scenarios. In the power splitting case (Fig.~\ref{fig:Fig5_Rho_Sec}), $P_{out}$ drops sharply after $\rho = 0.4$ and eventually reaches zero, indicating a significant improvement in UE throughput relative to Eve, as also seen in Fig.~\ref{fig:Fig5_Rho}. This results in a corresponding rise in $C_s$. Notably, the difference in $P_{out}$ and $C_s$ between 10 dB and 20 dB $P_{BS}$ is minimal, as the power allocation ratio between UE and Eve remains unchanged.
In contrast, Fig.~\ref{fig:Fig5_Beta_Sec} (antenna splitting case) shows a widening gap in both $P_{out}$ and $C_s$ as $\beta$ increases, primarily due to enhanced RIS gain. At lower $N$, the secrecy outage is higher, but it decreases rapidly with increasing $\beta$, since fewer RIS elements are reflecting toward Eve and its signal weakens with distance. The secrecy capacity improves accordingly, with a growing performance gap between $N = 32$ and $N = 64$, consistent with the trend observed in Fig.~\ref{fig:Fig5_Beta} for the throughput of UE.
\section{Conclusion and Future Directions}
\label{sec:Con}
RIS offers transformative potential for wireless communication through programmable radio environment control. However, its architecture introduces hardware-level vulnerabilities beyond traditional security models. This paper identified RIS-specific hardware threats, ranging from physical tampering and malicious reconfiguration to side-channel attacks, and highlighted their impact on confidentiality, integrity, and availability. These stealthy, trace-resistant threats underscore the urgent need for end-to-end security from design and manufacturing to deployment.

Future work will explore broader hardware-level attacks and practical defenses, including physical protection, secure reconfiguration, and trusted firmware. We also aim to investigate AI for real-time threat mitigation and RIS integration in multi-agent systems, where scalable, robust security is critical.
\section*{Acknowledgments}
This research was supported by the Sungkyunkwan University and the BK21 FOUR(Graduate School Innovation) funded by the Ministry of Education (MOE, Korea) and National Research Foundation of Korea (NRF), Institute of Information $\&$ communications Technology Planning $\&$ Evaluation (IITP) grant funded by the Korea government (MSIT) (RS-2024-00397216), the German Federal Ministry of Education and Research (BMBF) within the projects Open6GHub under grant number \{16KISK004\}, and under grant number \{03FHP106\}, as part of the “Career@BI” project within the FH Personal program.
\bibliographystyle{IEEEtran} 
\bibliography{reference.bib}
\end{document}